\documentstyle[floats,epsfig,aps]{revtex}

\newcommand{\beq}{\begin{equation}}
\newcommand{\eeq}[1]{\label{#1}\end{equation}}
\newcommand{\beqn}{\begin{eqnarray}}
\newcommand{\eeqn}[1]{\label{#1}\end{eqnarray}}
\def\nn{\nonumber}

\newcommand{\ba}{\begin{array}}
\newcommand{\ea}{\end{array}}
\newcommand{\Spin}{{\vec{S}}}

\begin{document}
\tolerance 50000 \preprint{
\begin{minipage}[t]{1.8in}
\rightline{La Plata-Th 00/xx} \rightline{}
\end{minipage}
}


\title{Non-perturbative effective field theory for two-leg antiferromagnetic spin ladders}

\author{D.C.\ Cabra$^{1,2}$, A.\ Dobry$^3$ and G.L.\ Rossini$^1$}
\address{
$^{1}$Departamento de F\'{\i}sica, Universidad Nacional de la
Plata, C.C.\ 67, (1900) La Plata, Argentina.\\ $^{2}$ Facultad de
Ingenier\'\i a, Universidad Nacional de Lomas de Zamora, Cno. de
Cintura y Juan XXIII,\\ (1832) Lomas de Zamora, Argentina.\\
$^3$Departamento de F\'{\i}sica, Facultad de Ciencias Exactas,
Ingenier\'\i a y Agrimensura and IFIR (CONICET-UNR)\\ Av.\
Pellegrini 250, 2000 Rosario, Argentina}
\date{\today}
\maketitle

\begin{abstract}
\begin{center}
\parbox{14cm}{
We study the long wavelength limit of a spin $1/2$ Heisenberg
antiferromagnetic two-leg ladder, treating the interchain coupling
in a non-perturbative way. We perform a mean field analysis and
then include exactly the fluctuations. This allows for a
discussion of the phase diagram of the system and provides an
effective field theory for the low energy excitations. The coset
fermionic Lagrangian obtained corresponds to a perturbed
$SU(4)_1/U(1)$ Conformal Field Theory (CFT). This effective theory
is naturally embedded in a $SU(2)_2 \times Z_2$ CFT, where
perturbations are easily identified in terms of conformal
operators in the two sectors. Crossed and zig-zag ladders are also
discussed using the same approach. }
\end{center}
\end{abstract}


\section{Introduction}

Antiferromagnetic Heisenberg spin ladders have been a subject of
central interest during the last years. These are intermediate
systems between the gapless critical spin $1/2$ Heisenberg chain
and the ordered spin $1/2$ 2D system relevant for undoped cuprate
superconductors. The simplest realization, {\it i.e.} the two-leg
ladder, shows a dramatic different excitation spectrum with
respect to the one of an isolated chain. It has a finite gap to
the first excitation and magnetic correlations are short ranged.
Several inorganic compounds have been recently synthesized and
modeled as Heisenberg ladders \cite{dagotto}. Exponential decay of
the low temperature magnetic susceptibility was the first signal
of the existence of a spin gap in two-leg ladder materials.
Neutron and optical measurements also manifest the presence of a
gap and are consistently described by a two-leg ladder model with
exchange integrals of the same order in the chains direction($J$)
and along the rungs ($J'$).

Theoretically, the existence of a gap was early predicted from
numerical exact diagonalization and  strong coupling perturbation
theory ($J/J'\ll 1$) \cite{Barnes}. More recently field
theoretical technics have been used to analyze the  excitation
spectrum in the weak coupling regime ($J'/J \ll 1$) \cite{TS,NST}.
These treatments give access to the whole low energy excitation
spectrum as well as to the dynamical susceptibilities, which are
essential to compare with experimental probes. The philosophy
underlying this study is the following: spin operators are
expressed in the well known bosonized representation of each chain
and the interchain coupling is treated as a small perturbation in
this representation. The applicability of these studies is then
valid in principle only in the weak coupling regime and its use in
the description of {\it e.g.} the experimentally realized two-leg
ladders in which $J' \sim J/2$ should be taken with some care. It is
therefore not clear up to which value of $\frac{J'}{J}$ the
results of \cite{TS,NST} are applicable, and it is important to develop
theoretical methods which could be used beyond the weak coupling
regime.

The picture that emerges from the weak coupling analysis leads to
a description in terms of triplet of massive Majorana fermions
and a singlet Majorana fermion with a different mass (which has
been estimated to be minus three times the triplet mass)
\cite{NST}. The only interactions between these fermions are marginal
current-current terms which have been argued to simply renormalize
their masses and velocities.
A question, that has been risen in recent studies of the Raman
scattering spectrum \cite{Orignac}, is whether marginal interactions can in fact be
disregarded. In particular, correlation
functions obtained disregarding marginal interactions apparently do not fit
experiments (see {\it e.g.} \cite{ATS}).

In this work we analyze the complete phase diagram of the two-leg
antiferromagnetic ladder. Our approach, first used here for spin
ladders, starts from a fermionic representation of the spin
operators in the functional integral framework, as introduced in
\cite{MA,IM,MF} for spin chains. With a simple Ansatz to the Mean
Field (MF) configurations we show that the system undergoes a
cross-over from a weak to a strong coupling regime at an
intermediate value of $J'/J$. We then introduce fluctuations
around MF and take them into account to all orders to construct
the low energy effective field theory.

The resulting theory corresponds to a coset Conformal Field Theory
(CFT) of symmetry $SU(4)_1/U(1)_{iso}$, perturbed by relevant
operators (of dimension $1$) and marginal operators (of dimension
$2$) arising from the single occupancy constraint as well as from
the amplitude fluctuations of the link fields introduced to
decouple the fermionic interactions. It should be noted that our
approach is based on the assumption that the local single
occupancy constraint can be implemented as a very last step, while
it is taken into account globally from the beginning. The
correctness of this procedure is not guaranteed from first
principles, but is supported a posteriori.

We show that the complete structure of these perturbations can be
retained and that they take a simple form in the language of
conformal embeddings. In particular, the marginal terms which
arise can be easily classified in the new language and their
effect can then be studied in a non-perturbative way. When the
relevant perturbations are expressed in the embedded $SU(2)_2
\times Z_2$ CFT language the spectrum is naturally separated in
the triplet and singlet of Majorana fermions. These results, which
are valid up to $J'/J \approx 8/\pi^2$, extend to finite coupling
the weak coupling study of \cite{NST}. It should be stressed that
recent estimates of the ratio of exchange constants lead to values
of $J'/J$ around $1/2$ in several cuprate materials \cite{FS}.

In order to illustrate the generality and ease of use of our
approach, it is then applied to the so-called crossed ladders and
zig-zag ladders. Phase diagrams and low energy theories are
obtained in the region containing the weak coupling limit; further
analysis and details will be considered elsewhere.

The paper is organized as follows. In Section II we introduce the
model, present Hubbard-Stratonovich decoupling technics and
perform a MF analysis, discussing the resulting phase diagram. In
Section III we construct the low energy effective field theory:
our theory contains four Dirac fermionic species corresponding to
the spin and band indices of the ladder. In Section IV we show
that the theory has a natural relation to $SU(2)_2 \times Z_2$ CFT
through conformal embedding (the last part arises from the two
electronic bands). In Section V we briefly report results on
crossed and zig-zag ladders. Finally, in Section VI the
conclusions and possible further developments of our method are
given.

\section{Mean field analysis}

We consider the Heisenberg Hamiltonian for a two-leg spin $1/2$
ladder,
\beq
H=\sum_{n=1}^N  \sum_{l=1}^2 \left(J
\Spin^{(l)}_{n} . \Spin^{(l)}_{n+1} + \frac{J'}{2} \Spin^{(l)}_{n} .
\Spin^{(l+1)}_{n} \right)
\eeq{H}
where $N$ is the number of sites along the chains, $J>0$ and
$J'>0$ are the couplings between adjacent spins along the legs and
rungs respectively. For mathematical convenience we assume
periodic boundary conditions (P.B.C.) in both directions (notice
that the Hamiltonian is suitable written for arbitrary $n$-leg
ladders; in the present case the physical coupling along the rungs
is in effect $J'$).

The spin variables can be represented in terms of fermionic
operators with spin $c^{(l)}_{n,\alpha}$ as
\beq
\Spin^{(l)}_{n} = c^{\dagger (l)}_{n,\alpha}
\frac{{\vec{\sigma}}_{\alpha\beta}}{2} c^{(l)}_{n,\beta},
\eeq{spin-rep}
where $\vec{\sigma}$ are
Pauli matrices, together with a local constraint that ensures one
spin per site, imposed on the physical states by
\beq
c^{\dagger(l)}_{n,\alpha}c^{(l)}_{n,\alpha} \vert phys \rangle
= \vert phys \rangle \ .
\eeq{constraint}
Throughout this paper we will not use the summation convention
neither for site nor leg indices; repeated spin (Greek) indices
are summed.

Using (\ref{spin-rep}) and (\ref{constraint})
the Hamiltonian (\ref{H}) can be rewritten as
\beq
H= -\sum_{n=1}^N  \sum_{l=1}^2 \left(\frac{J}{2} c^{\dagger
(l)}_{n,\alpha} c^{(l)}_{n+1,\alpha} c^{\dagger (l)}_{n+1,\beta}
c^{(l)}_{n,\beta}  + \frac{J'}{4} c^{\dagger (l)}_{n,\alpha}
c^{(l+1)}_{n,\alpha} c^{\dagger (l+1)}_{n,\beta} c^{(l)}_{n,\beta}
\right) + C
\eeq{H-quartic}
where $C= -N/2$ is an irrelevant constant term.

We now trade (\ref{H-quartic}) for a quadratic Hamiltonian via a
Hubbard-Stratonovich transformation, at the usual price of
introducing auxiliary fields $B^{(l)}_n$
associated to terms containing  $c^{\dagger (l)}_{n+1,\alpha}
c^{(l)}_{n,\alpha}$ and $B'_{n}$ associated to terms containing
$c^{\dagger (l+1)}_{n,\beta} c^{(l)}_{n,\beta}$. It is natural to
interpret $B^{(l)}_{n}$ as localized on the leg $(l)$ links between
sites $n$ and $n+1$ and $B'_{n}$ as localized on the rung links.
After the transformation the Hamiltonian reads
\beqn
H &=& \frac{J}{2} \sum_{n=1}^N \sum_{l=1}^2 \left(
B^{(l)}_{n}c^{\dagger (l)}_{n,\alpha} c^{(l)}_{n+1,\alpha} +
B^{\dagger (l)}_{n} c^{\dagger (l)}_{n+1,\beta} c^{(l)}_{n,\beta}
+ B^{\dagger (l)}_{n}B^{(l)}_{n}  \right) \nn \\ &+&\frac{J'}{4}
\sum_{n=1}^N \sum_{l=1}^2 \left( B'_{n}c^{\dagger (l)}_{n,\alpha}
c^{(l+1)}_{n,\alpha} + B^{\dagger '}_{n} c^{\dagger
(l+1)}_{n,\beta} c^{(l)}_{n,\beta} + B^{\dagger '}_{n}B^{'}_{n}
\right)
\eeqn{HcB}

As we look for a low energy effective theory, we treat the $B$
variables in a long wave approximation. To this end, we
parameterize these fields in terms of real MF values
$(B_0,B^{'}_0)$ and fluctuations

\beq B^{(l)}_{n} = B_0 \exp (i a A^{(l)}_n + a R^{(l)}_n), \ \ \
B^{'}_{n} =  B^{'}_0 \exp (i a A^{'}_n + a R^{'}_n)\ . \eeq{mf}
Notice that we have included  both phase and amplitude
fluctuations, which will play important different r\^oles in the
following. For this reason, we explicitly distinguish the
Hermitean $(R^{(l)}_n , R^{'(l)}_n)$, and anti-Hermitean
$(iA^{(l)}_n , iA^{'(l)}_n)$ parts of the fluctuation fields.
The expression for $B^{'}_{n}$ will be eventually modified when
$B^{'}_0=0$ (see eq.\ (\ref{Bexpansion})).

As a first step, we perform the MF evaluation of the Hamiltonian
(\ref{HcB}) by setting the fluctuations to zero. The resulting MF
Hamiltonian is then a tight-binding model for two coupled chains,
\beqn
H_{mf}& =& -t \sum_{n=1}^N \sum_{l=1}^2 \left(c^{\dagger
(l)}_{n,\alpha} c^{(l)}_{n+1,\alpha} + c^{\dagger (l)}_{n+1,\beta}
c^{(l)}_{n,\beta}\right) - 2 t' \sum_{n=1}^N \left( c^{\dagger
(1)}_{n,\alpha} c^{(2)}_{n,\alpha}+ c^{\dagger (2)}_{n,\beta}
c^{(1)}_{n,\beta} \right)\nonumber \\ & &
 + \frac{4N}{J}t^2 +  \frac{8N}{J'}{t'}^2,
\eeqn{HmfComp}
where
\beq
t = - \frac{J B_0}{2} , \ \ \
t' = -\frac{J'B^{'}_0}{4} \ .
\eeq{ts}

The coupled tight-binding model is easily diagonalized by means of
a double Fourier transform. We first decouple two bands by means
of

\beq c^{(1)}_{n,\alpha} = \frac{1}{\sqrt{2}} (c^{(+)}_{n,\alpha}-
c^{(-)}_{n,\alpha}), \eeq{c1a}
\beq c^{(2)}_{n,\alpha} = \frac{1}{\sqrt{2}} (c^{(+)}_{n,\alpha}+
c^{(-)}_{n,\alpha}) \eeq{c2a}
and then introduce pseudo-momentum operators $d^{(+)}, d^{(-)}$ by
\beq c^{(+)}_{n,\alpha} = \frac{1}{\sqrt{2N}} \sum_{m=1}^N
d^{(+)}_{m,\alpha}\exp(-i\frac{2\pi mn}{N}), \eeq{Fourier1}
\beq
 c^{(-)}_{n,\alpha} = \frac{1}{\sqrt{2N}} \sum_{m=1}^N
 d^{(-)}_{m,\alpha}\exp(-i\frac{2\pi mn}{N}),
\eeq{Fourier2}
in terms of which the Hamiltonian reads
\beqn
H_{mf}&=& -
\sum_{m=1}^N(2t\cos(\frac{2\pi}{N}m)+2t')d^{\dagger
(+)}_{m,\alpha} d^{(+)}_{m,\alpha} - \nonumber \\ & &
\sum_{m=1}^N(2t\cos(\frac{2\pi}{N}m)-2t')d^{\dagger
(-)}_{m,\alpha} d^{(-)}_{m,\alpha}  + \frac{4N}{J}t^2 +
\frac{8N}{J'}{t'}^2.
\eeqn{Hmfdiag}

This expression clearly represents a decoupled two-band
tight-binding model.

The constraint (\ref{constraint}), meaning one electron per site,
forces the system to be exactly at half filling. Low-energy
excitations are then achieved by creating holes just below the
Fermi surface and creating electrons just above it\cite{LesH}.
Notice that this can be done only if
\beq |t'| < |t|, \eeq{condition}
that is when the Fermi level crosses both bands. If this condition
is not satisfied, the system presents a finite energy gap to spin
excitations.

The actual values of $t$ and $t'$ are determined by minimizing the
energy of (\ref{Hmfdiag}). In order
to perform this evaluation we introduce a lattice spacing $a$ and
a position coordinate $x=na$ ($x \in [0,L=Na]$); the appropriate
pseudo-momentum coordinate is $k= m2\pi /(Na)$ ($ k\in [ -\pi/a,
\pi/a]$). The mean-field Hamiltonian then reads
\beqn
H_{mf} & = & -\frac{L}{2\pi}\int_{-\pi/a}^{\pi/a}(2t\cos(k
a)+2t')d^{\dagger (+)}_{\alpha}(k) d^{(+)}_{\alpha}(k) dk -
\nonumber \\ & & \frac{L}{2\pi}\int_{-\pi/a}^{\pi/a}(2t\cos(k
a)-2t')d^{\dagger (-)}_{\alpha}(k) d^{(-)}_{\alpha}(k) dk
\nonumber \\ & &
 + \frac{4L}{aJ}t^2 +\frac{8L}{aJ'}{t'}^2,
\eeqn {Hmfcontinuo}
from which we can read the dispersion relations for each band,
sketched in Fig.\ 1,
\beq \epsilon^{(+)}(k)= - 2t\cos(k a)-2t', \eeq{energy+}
\beq \epsilon^{(-)}(k)= - 2t\cos(k a)+2t'. \eeq{energy-}

\begin{figure}[htp]
\begin{center}
\epsfig{file=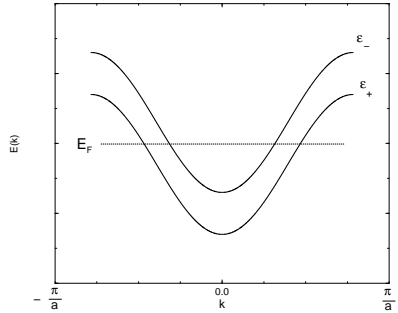,width=5cm,angle=-90}
\end{center}
\vskip .5 truecm

\caption{Dispersion relation for the two-band tight-binding model}
\end{figure}

The Fermi momentum for each band is defined through the equation
\beq \epsilon^{(+)}(k_F^{(+)})=\epsilon^{(-)}(k_F^{(-)}).
\eeq{kFs}
The local constraint in eq.\ (\ref{constraint}) leads to the
global constraint $ N^{(+)}+N^{(-)}=2N$,
where $N^{(\pm)}$ is the occupation number operator for each band.
Besides, $N^{(\pm)}= \frac{2Na}{\pi}k^{(\pm)}_F$.
We thus obtain
\beq
\cos(k_F^{(-)}a)=\frac{t'}{t} = \frac{J'B'_0}{2 JB_0}
\eeq{kF-}
\beq
k_F^{(+)}= \pi /a -k_F^{(-)},
\eeq{kF+}
these implying
\beq
\epsilon^{(+)}(k_F^{(+)})=\epsilon^{(-)}(k_F^{(-)})=0.
\eeq{EF=0}

The values of $t$ and $t'$  are now determined by minimizing the
energy of (\ref{Hmfcontinuo}) under half-filling conditions,
\beqn E_{mf} & = & -\frac{L}{2\pi}\int_{-k_F^{(+)}}^{k_F^{(+)}}
(2t\cos(k a)+2t')\, 2\,  dk -
\frac{L}{2\pi}\int_{-k_F^{(-)}}^{k_F^{(-)}}(2t\cos(k a)-2t')\, 2\,
dk \nonumber \\ & &
 +\frac{4L}{aJ}t^2 +\frac{8L}{aJ'}{t'}^2 .
\eeqn{Emfgral}
Notice that $t<0$ just inverts the cosine curves, translating the
Brillouin zone considered in $\pi/a$, and $t'<0$ would just trade
the roles of the two bands. Then, the relevant sector in the
$t,t'$ plane is $t\geq 0$ and $t'\geq 0$. In this sector, the
expression for the energy is
\beq
\frac{a\pi}{L}E_{mf}= -8 t \sin(k_F^{(-)}a)-4 t'(\pi -
2k_F^{(-)}a) +  \frac{4\pi}{J}t^2 +\frac{8\pi}{J'}{t'}^2,
~~~~~~~t'<t,\eeq{Emf1}
and
\beq
\frac{a\pi}{L}E_{mf}= -4\pi t' +  \frac{4\pi}{J}t^2
+\frac{8\pi}{J'}{t'}^2, ~~~~~~~~~~~~~~~~~~~~~~~~~~~~~~~~~~~~~~
t'>t.
\eeq{Emf2}

The analysis of the above equations shows that, for $J'< 8/\pi^2 ~
J$, the MF configuration depicts two bands which coincide with
those corresponding to two decoupled chains ($t=J/\pi$, $t'=0$),
with Fermi momentum $k_F=k_F^{(-)}=k_F^{(+)}=\pi/(2a)$. Notice
that the condition in eq.\ (\ref{condition}) holds and the
linearization procedure around this minimum is valid.

On the contrary, for $ J' > \frac{8}{\pi^2} J$, we find that the
global energy minimum corresponds to the point $t=0$, $t'=J'/4$
where the condition in eq.\ (\ref{condition}) does not hold
(notice that there is still another local minimum while $J'< 2
J$). The system in this configuration, which describes the strong
coupling phase, presents a finite energy gap to spin excitations.

In the following two sections we will  explore the $J'< 8/\pi^2 ~ J$
region.

\section{Fluctuations and constraints: the $SU(4)_1/U(1)$ coset theory}

In this section we take the continuum limit of the MF Hamiltonian
in eq.\ (\ref{Hmfdiag}) and then include fluctuations around MF
and constraints in eq.\ (\ref{constraint}). The outcome of
this procedure is a perturbed  $SU(4)_1/U(1)$ coset theory.

\subsection{Low-energy linearization in the thermodynamical continuum
limit}

In the region we consider ($J'< 8/\pi^2 J$) the mean-field
dispersion relation consists of two coinciding bands of amplitude
$2J/\pi$. Linearization of low-energy excitations can be done
around $k_F = \pi/(2a)$ in the usual way. The bandwidth will limit
the validity of the resulting effective Field Theory to energies
much smaller than $J$, independently of $J'$.

Low-energy excitations in the thermodynamical continuum limit of
the tight-binding model at half filling can be linearized in terms
of Dirac fermions \cite{LesH}. Fermionic position space operators
$c^{(\pm)}$ for each band are readily written in terms of Dirac
fermions $\psi^{(\pm)}(x)$ as
\beq c^{(+)}_{n,\alpha}= \sqrt{a}(\exp(-ik_F
x)\psi^{(+)}_{R,\alpha}(x) + \exp(ik_F
x)\psi^{(+)}_{L,\alpha}(x)), \eeq{psi+}
\beq c^{(-)}_{n,\alpha}= \sqrt{a}(\exp(-ik_F
x)\psi^{(-)}_{R,\alpha}(x) + \exp(ik_F
x)\psi^{(-)}_{L,\alpha}(x)). \eeq{psi-}
Here $\psi^{(+)}_{R,\alpha}$ and $\psi^{(+)}_{L,\alpha}$ stand for
the right and left components of a Dirac spinor
$\Psi^{(+)}_{\alpha}$ and so on. Dirac gamma matrices are taken as
$\gamma_0=\sigma_1$, $\gamma_1=\sigma_2$. Notice that there is a
total of four Dirac fermion species; using the notation
$\alpha=\uparrow , \downarrow$ they are
$(+,\uparrow),(-,\uparrow),(+,\downarrow), (-,\downarrow)$, which
will be respectively denoted $\Psi_i(x)$, with $i=1,2,3,4$.
Summation over repeated fermion species indices will be
understood.

We arrive then at the linearized MF Hamiltonian
\beq H_{mf}= v_F \int dx  \bar{\Psi}_i(x) \gamma_1\partial_x
\Psi_i(x), \eeq{Hmfpsi}
where
\beq v_F= 2 t a \sin(k_F a)=2 J a/\pi \eeq{vFermi}
is the Fermi velocity, and $\bar{\Psi}_i= \Psi^{\dagger}_i
\gamma_0$. All four $\Psi$-fields have the same Fermi velocity,
then the model, up to this point, possesses a manifest $U(4)$
symmetry.

\subsection{Fluctuations around mean field}

We include now the fluctuation fields $B^{(1)}_n, B^{(2)}_n
,B'_n$. As we look for the continuum limit of the Hamiltonian
(\ref{HcB}), we will keep only relevant powers in $a$, as compared
with eq.\ (\ref{Hmfpsi}).

In order to keep track of $a$ orders, it is useful to make
explicit the order $a$ contribution of fermion bilinears by
defining
\beq \ba{l} z^{(1)}_n=a^{-1} c^{\dagger (1)}_{n,\alpha}
c^{(1)}_{n+1,\alpha},
\\
z^{(2)}_n=a^{-1}c^{\dagger (2)}_{n,\alpha} c^{(2)}_{n+1,\alpha},
 \\
z^{(3)}_n=a^{-1}c^{\dagger (1)}_{n,\alpha} c^{(2)}_{n,\alpha}, \ea
\eeq{z123}
so that the leading order for each $z$ is $a^0$. Notice that
$z^{(1)}$ and $z^{(2)}$ still have to be expanded, as
$(n+1)a=x+a$; the only relevant term in this expansion is that
linear in $a$, containing first derivatives of $\psi$-fields. Our
notation will be
\beq z^{(i)}_n= w^{(i)}_n + av^{(i)}_n \eeq{wv}
(notice that $v^{(3)}_n=0$).

The relevant expansions for the $B$-fields, taking into account
that the MF value of ${B'}_0$ vanishes, are
\beq \ba{l} B^{(l)}_n = B_0+ i a B_0 A^{(l)}_n + a B_0 R^{(l)}_n +
O(a^2), \ \ \ \ (l=1,2)
\\
{B'}_n = i a A'_n + a  R'_n + O(a^2). \ea \eeq{Bexpansion}
In particular, the terms quadratic in $B$ must be expanded as
\beq \ba{l} B^{\dagger(l)}_n B^{(l)}_n = B_0^2 + 2 a B_0^2
R^{(l)}_n + 2 a^2 B_0^2 {R^{(l)}}^2_n, \ \ \ \ (l=1,2)
\\
B^{'\dagger}_n B'_n = a^2 {A'}^2_n + a^2 {R'}^2_n . \ea
\eeq{B2expansion}

Using all of these, and making explicit the sums over $l=1,2$, the
effective low energy Hamiltonian for (\ref{HcB}) is written as
\beq H_{eff}= H^{(1)}+ H^{(2)}+H^{(3)} + O(a^3) \eeq{Hsuma}
where
\beqn H^{(1)} &=& \frac{J}{2} N B_0^2+ \frac{J}{2} \sum_{n=1}^N a
B_0 (z^{(1)}_n + z^{\dagger(1)}_n) + \frac{J}{2}\sum_{n=1}^N i a^2
B_0 A^{(1)}_n (w^{(1)}_n -w^{\dagger(1)}_n) + \nonumber \\ & &
\frac{J}{2}\sum_{n=1}^N a (a B_0 R^{(1)}_n (w^{(1)}_n +
w^{\dagger(1)}_n) +2 B^2_0 R^{(1)}_n + 2 a B^2_0 {R^{(1)}}^2_n),
\eeqn{H1}
\beqn H^{(2)} &=& \frac{J}{2} N B_0^2+ \frac{J}{2} \sum_{n=1}^N a
B_0 (z^{(2)}_n +z^{\dagger(2)}_n) + \frac{J}{2}\sum_{n=1}^N i a^2
B_0 A^{(2)}_n (w^{(2)}_n -w^{\dagger(2)}_n) + \nonumber \\ & &
\frac{J}{2}\sum_{n=1}^N a (a B_0 R^{(2)}_n (w^{(2)}_n
+w^{\dagger(2)}_n ) +2 B^2_0 R^{(2)}_n + 2 a B^2_0 {R^{(2)}}^2_n),
\eeqn{H2}
\beqn
H^{(3)} &=& \frac{J'}{2} \sum_{n=1}^N i a^2 [ A'_n (w^{(3)}_n
-w^{\dagger(3)}_n) + {A'}^2_n] + \nonumber \\ & & \frac{J'}{2} \sum_{n=1}^N
a^2 [R'_n (w^{(3)}_n + w^{\dagger(3)}_n) + {R'}^2_n]
\eeqn{H3}

The main things to notice here are:

- there are irrelevant (divergent) constant terms. This is
expected from the combination of Hubbard-Stratonovich and MF
techniques.

- the terms without fluctuations in $H^{(1)}$ and $H^{(2)}$
provide the two decoupled chains MF results discussed in the
previous section.

- The $A^{(1)}$ and $A^{(2)}$ fields act as Lagrange multipliers;
their total contribution to the effective action in the continuum
limit reduces to a term
\beq
-i \frac{v_F}{2} \int dx \left( (\bar{\Psi}_i(x) \gamma_1
\Psi_i(x)) (A^{(1)}(x) + A^{(2)}(x))+ (\bar{\Psi}_i(x) \gamma_1
(\sigma_1 \otimes 1 )_{ij}\Psi_j(x)) (A^{(1)}(x) - A^{(2)}(x))
\right)\ .
\eeq{gauged}
In the notation of eq.\ (\ref{gauged}) the first matrix
($\sigma_1$) refers to isospin indices $(+),(-)$, while the second
one ($1$) refers to spin indices $\uparrow,\downarrow$.

 - The presence of a quadratic term in the ${A'}$ field, with proper
sign, allows for a trivial Gaussian integration. The same can be
done with the $R$ fields. These of course brings back the original
spin-spin rung interactions. In the present scheme their
contribution includes quadratic terms in the $c$ operators, that
lead to a redefinition of the Fermi velocity $v_F \rightarrow
v_F/2$, and quartic perturbations that can be arranged as

\beq - \frac{Ja^2}{16}\sum_{n=1}^N  \left( (w^{(1)}_n +
w^{\dagger(1)}_n )^2 + (w^{(2)}_n + w^{\dagger(2)}_n )^2\right) -
\frac{J' a^2}{2} \sum_{n=1}^N  w^{(3)}_n w^{\dagger (3)}_n . \eeq{PF}
The continuum form of these quartic perturbations in terms of
Dirac fermions is lengthy. We will write them down below, after
introducing a convenient notation.

We notice that, for $J'< 8/\pi^2 ~J$, our approach leads to a
description of the system which is the same as the one obtained in
perturbative treatments, in principle valid for $J' \ll J$
\cite{NST,H}. In particular, the first two terms in (\ref{PF})
give rise to the well known marginally irrelevant perturbation
terms in the individual chains. However, our approach does not
rely on any perturbative treatment of $J'$ and in particular
allows for the determination of the phase diagram of the system,
{\it i.e.} it predicts a critical value of the ratio $J'/J$ which
separates the two different regimes in the two-leg ladder. The
situation is depicted in Fig.\ 2.

\begin{figure}[hpt]
\begin{center}
\epsfig{file=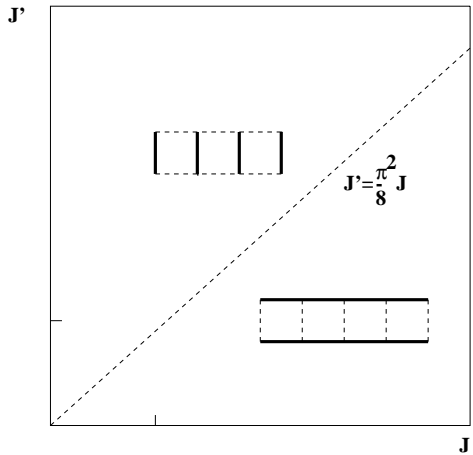,width=6cm,angle=-90}
\end{center}
\vskip .5 truecm

\caption{Phase diagram of the spin ladder. Bold bonds correspond
to non zero links in the MF approximation. }
\end{figure}

Moreover, we show in the next section that the weak coupling
structure unraveled in \cite{NST} arises naturally within our
approach.

\subsection{Constraints}

In this section we express the constraints (\ref{constraint}) in
terms of the linearized fermion fields and discuss how to
implement them in the evaluation of the partition function for the
spin ladder.

In the continuum limit the constraint on the occupation number at
each site $(l),n$ separates in four parts, corresponding to
oscillating and non-oscillating terms associated to each band.
They read:

\beq
\bar{\Psi}_i \gamma_0 \Psi_i =constant
\eeq{a0_diagonal}
which, implemented through a Lagrange multiplier $a_0$, provides
the time component of a gauge field implementing a diagonal $U(1)$
coset constraint,  $a_{\mu}=(a_0,A^{(1)}+A^{(2)})$;

\beq
\bar{\Psi}_i \gamma_0 (\sigma_1 \otimes 1 \!\! 1)_{ij} \Psi_j =0
\eeq{b0_isospin}
which, implemented through a Lagrange multiplier $b_0$, provides
the time component of a second gauge field implementing the
isospin $U(1)$ coset constraint, $b_{\mu}=(b_0,A^{(1)}-A{(2)})$.
These last two constraints

\beq
\bar{\Psi}_i  \Psi_i=0 ;
\eeq{lambda_diagonal}

\beq
\bar{\Psi}_i (\sigma_1 \otimes 1 \!\! 1)_{ij} \Psi_j=0 \ ,
\eeq{lambda_isospin}
lead to marginally irrelevant quartic
perturbation terms when implemented through
\beq \delta({\cal O}(x))\propto \lim_{\eta \rightarrow \infty}
e^{-\eta{\cal O}^{\dagger} (x){\cal O}(x)}, \eeq{delta}
just as in the case of decoupled chains.

To conclude this section, we collect all the terms in the
effective low energy Hamiltonian which finally reads
\beq
 H_{eff}= \frac{aJ}{\pi} \int dx ~ \bar{\Psi}_i(x)
\left(\left(\gamma_1\partial_x - i \gamma_{\mu} a_\mu \right)\delta_{ij}
-i \gamma_{\mu}(\sigma_1\otimes 1 \!\! 1)_{ij} b_\mu\right) \Psi_j(x) +
\Delta H_{eff}\ ,
\eeq{coset}
where $\Delta H_{eff}$ includes quartic terms in fermionic fields,
which arise from (\ref{PF}) and
(\ref{lambda_diagonal}),(\ref{lambda_isospin}).

Clearly, the unperturbed theory posses a $U(4)$ symmetry which is gauged by a
diagonal $U(1)$ field ($a_\mu$) and an isospin $U(1)$ field
($b_\mu$) which leads to the coset
\beq
\frac{U(4)}{U(1)_{diag}
\times U(1)_{iso}}=\frac{SU(4)_1}{U(1)_{iso}},
\eeq{U4}
where $SU(4)_1$ stands for the level $k=1$ WZW theory \cite{Wi,KZ}.

Before displaying the explicit expression
for the perturbations it is worth discussing in more detail the coset structure of the
quadratic part of the Hamiltonian.

\section{$SU(2)_2 \times Z_2$ embedding, the perturbations in a
new language}

As it is known, the coset CFT $SU(4)_1$ can be alternatively
described through the embedding  \cite{BS}
\beq
 SU(4)_1= SU(2)_2 \times SU(2)_2 \ .
\eeq{SU2xSU2}
The conformal central charges of the two theories coincide and
primary fields in the $SU(4)_1$ can be written in terms of
primaries in the two $SU(2)_2$ sectors. This will presently prove to be
useful in the treatment of the perturbations. The
different $SU(2)_2$ sectors in this embedding are naturally
identified in eq.\ (\ref{coset}) as the spin and isospin sectors,
in virtue of the $\sigma_1 \otimes 1 \!\! 1$ non-diagonal
structure. Moreover, in this language the second Lagrange
multiplier $b_\mu$ gauges a $U(1)$ subgroup of the isospin
$SU(2)_2$ sector giving rise to
\beq
SU(2)_2^{spin} \times
SU(2)_2^{isospin}/U(1)^{isospin} = SU(2)_2^{spin} \times Z_2.
\eeq{SU2xSU2/U(1)}
The last factor has been identified in \cite{NST} from the $Z_2$ structure of a two
chain system.

All of this is most easily shown in the bosonized version of the
coset CFT. To this end we write fermion bilinears as \cite{Wi,NS}
\beq \psi^{\dagger ~i}_R \psi^{\bar \jmath}_L = M
\Phi_{\Box}^{i{\bar \jmath}} \eeq{bosrule}
where $M$ is a renormalization constant and we have introduced bar
indices in order to distinguish components transforming in the
right and left fundamental representations of $SU(4)_1$. The
$\Box$ subindex indicates the fundamental representation in the
standard Young tableaux notation. In identifying the two $SU(2)_2$
sectors we find useful to keep the original spin and isospin
(band) indices, writing
\beq \Phi_{\Box}^{i{\bar \jmath}}=\Phi_{\Box}^{a \alpha, \bar b
\bar \beta} \eeq{oldindices}
where we now use $a, \bar b$ for $(+),(-)$.

This field $\Phi_{\Box}^{i{\bar \jmath}} $ has scaling dimension $3/4$ and its components can be
written in terms of products of the components of the fields in
the fundamental representations of the two $SU(2)_2$ sectors as
\beq
\Phi_{\Box}^{a \alpha, \bar b \bar \beta} =
\phi_{\Box}^{a,\bar b}{\phi}_{\Box}^{'\alpha,\bar \beta}
\eeq{id1}
where $\phi_\Box$ and ${\phi'}_\Box$ are the primary fields in the
fundamental (spin 1/2) representation of the two $SU(2)_2$ isospin
and spin
sectors respectively. These fields have scaling dimension $3/8$ so the product
has the right dimension $3/4$ and moreover, correlation functions
of the fields on both sides coincide.

The other primary field in the $SU(4)_1$ CFT is the one
transforming in the antisymmetric ($6 \times \bar 6$)
representation, which in the Young tableaux notation should read
$\Phi_{\stackrel{\textstyle{\small\Box}}{\Box }}$. It is built
up from the antisymmetric product of two fields in the fundamental
representation
\beq
\Phi_{\stackrel{\textstyle{\small\Box}}{\Box }}= {\cal A}
\left( \Phi_{\Box}\Phi_{\Box} \right)
\eeq{asym}
This field has scaling dimension $1$ and can be mapped into
$SU(2)_2$ fields as
\beq
\Phi_{\stackrel{\textstyle{\small\Box}}{\Box}}^{[(a_1\alpha_1),(a_2,\alpha_2)],
[(\bar b_1\bar\beta_1),(\bar b_2,\bar\beta_2)] }
=\phi_{\Box \! \Box}^{\{a_1 a_2\},\{\bar b_1 \bar b_2\}}
\epsilon^{\alpha_1\alpha_2} \epsilon^{\bar\beta_1 \bar \beta_2}  +
\epsilon^{a_1 a_2}\epsilon^{\bar b_1 \bar b_2} \phi_{\Box
\! \Box}^{'\{\alpha_1\alpha_2\} \{\bar\beta_1 \bar \beta_2\}}
\eeq{sim}
where $\phi_{\Box \! \Box},\phi_{\Box \! \Box}' $ are the primary
fields in the symmetric (spin 1) representations of the two
$SU(2)_2$ sectors, which have the correct scaling dimension 1. In
eq.\ (\ref{sim}) we have used the symbols $\{, \}$ and $[,]$ to
indicate respectively symmetrization and antisymmetrization of
indices.

We are now ready to analyze the different perturbation terms in
$\Delta H_{eff}$. First of all, contribution coming from
intrachain couplings and constraints are known to be marginally
irrelevant, just as in the case of decoupled chains \cite{IM}.

The interchain perturbation terms in $\Delta H_{eff}$
(those arising from the last term in (\ref{PF})) can be separated into
two groups according to their scaling dimensions: there are terms which correspond to
relevant operators (scaling dimension 1) which can be identified
with certain linear combination of the components of the primary
(\ref{asym}) in the coset theory (\ref{coset}) and current-current
terms, which have scaling dimension 2 and are hence marginal.

More precisely, for the relevant part we can write
\beq
\mbox{relevant perturbations} = -\lambda \int dx \left(
\mbox{Tr}\left(A \Phi_{\stackrel{\textstyle{\small\Box}}{\Box
}}\right) + H.c.\ \right) \ ,
\eeq{pertfull}
where $\lambda \propto J'$ and $A$ is given by
\beq A=\left(\begin{array}{cccccc}
-1&0&0&0&0&0\\ 0&1/2&0&0&-1/2&0\\ 0&0&-1/2&1/2&0&0\\
0&0&1/2&-1/2&0&0\\ 0&-1/2&0&0&1/2&0\\ 0&0&0&0&0&-1
\end{array}\right)
\eeq{matrix}
(see the Appendix for details).
Using the identifications described above and after some
straightforward algebra we can readily identify the perturbation
terms (\ref{pertfull}) in the embedding theory as

\beq \mbox{relevant perturbations} = -\lambda \int dx \mbox{Tr}
\left(\phi_{\Box \! \Box}+H.c. \right) + \frac{\lambda}{2}\int dx
\mbox{Tr} \left(\phi'_{\Box} \sigma_1 \phi_{\Box}^{' \dagger}
\sigma_1 +H.c. \right)
\eeq{finalpert}
To analyze the effect of these perturbation terms it is convenient
to reformulate the $SU(2)_2$ WZW sector in terms of three
decoupled Majorana fermions, and in this new language it is easy
to see that the first term gives a mass to all three Majorana
fields \cite{FZ}. The second one is simply the energy operator of
the remaining Majorana sector \cite{KZ,CM}. Being all
perturbations of dimension $1$ we see that the gap opens linearly
with the interchain coupling as predicted from the weak coupling
limit \cite{TS,NST}. Note the different sign in the masses of the
two sectors, also in agreement with the weak coupling analysis.

As for the current-current terms, they correspond to marginal
perturbations and can be written as
\beqn
\mbox{marginal perturbations} = -\frac{a J'}{8}\int dx
\left( - \psi^{\dagger}_L ((\sigma_3+i\sigma_2)\otimes 1 \!\! 1)
\psi_L \psi^{\dagger}_L ((\sigma_3-i\sigma_2)\otimes 1 \!\! 1)
\psi_L - R \leftrightarrow L \right)\nonumber\\ -\frac{a
J'}{8}\int dx \left(  \psi^{\dagger}_L
((\sigma_3+i\sigma_2)\otimes 1 \!\! 1) \psi_L \psi^{\dagger}_R
((\sigma_3-i\sigma_2)\otimes 1 \!\! 1) \psi_R + R \leftrightarrow
L \right)\nonumber\\ -\frac{a J'}{8}\int dx \left(
-\psi^{\dagger}_L (1 \!\! 1 \otimes 1 \!\!1) \psi_L
\psi^{\dagger}_R (1 \!\! 1 \otimes 1 \!\!1)\psi_R -
\psi^{\dagger}_L (1 \!\! 1 \otimes  \vec{\sigma}) \psi_L  .
\psi^{\dagger}_R (1 \!\! 1 \otimes  \vec{\sigma})\psi_R
\right)\nonumber\\ - \frac{a J'}{8}\int dx \left(\psi^{\dagger}_L
(\sigma_1 \otimes 1 \!\! 1) \psi_L \psi^{\dagger}_R (\sigma_1
\otimes  1 \!\! 1)\psi_R + \psi^{\dagger}_L (\sigma_1 \otimes
\vec{\sigma}) \psi_L . \psi^{\dagger}_R (\sigma_1 \otimes
\vec{\sigma}) \psi_R \right)
\eeqn{merda}
The first two terms (first line) renormalize the Fermi velocity of
the Majorana (isospin) sector, while the third and fourth (second
line) correspond to marginal forward-scattering terms in the same
sector. The fifth and seventh terms are effectively zero due to
the constraints on the two corresponding $U(1)$ currents and the
sixth term correspond to the marginal forward-scattering terms in
the spin sector. The very last one mixes spin and isospin sectors.
This last contribution is nevertheless marginal, so it does not
change the low energy physics which in the present case is
dominated by the relevant perturbations already discussed. Its
effect could be important in the analysis of {\it e.g.} zig-zag
ladders where the relevant perturbations are wiped out, as we show
in the next Section, and only marginal interactions play a r\^ole
\cite{WA,AS,NGE,CHP}.

It can be easily shown that the marginal terms which are present
on each separate chain, written in the present language correspond
to the sixth and eight terms in the above expression. Due to the
fact that these terms correspond to marginally irrelevant couplings
and that they form a closed algebra, they will have no effect in
the low energy dynamics whatsoever. After having observed that,
one can see that the effective theory consists of two sectors
which are decoupled from each other.


\section{Other structures: crossed and zig-zag ladders}

In this section we will extend our previous analysis to more
general situations, which are not only of academic interest, but
are relevant in the analysis of real materials. These more general
situations arise when other (diagonal) couplings between spins in
neighbouring chains are not negligible. The two
structures that we analyze now are the so-called crossed ladders
\cite{WKO,W}, in which couplings along the two diagonals are
added, and zig-zag ladders in which only one diagonal coupling is
added \cite{WA,AS,NGE,CHP}. Another potential application of the present
formalism would be the study of the interplay between interchain
coupling and dimerization along the legs \cite{Sierra,CG,WN}.

\vspace{.5cm}

\noindent {\it i) Crossed ladders}

\vspace{.5cm}

We consider a Heisenberg Hamiltonian given by

\beq H=\sum_{n=1}^N  \sum_{l=1}^2 J \Spin^{(l)}_{n} .
\Spin^{(l)}_{n+1} + J' \sum_{n=1}^N \Spin^{(1)}_{n} .
\Spin^{(2)}_{n} + J_{\times} \sum_{n=1}^N \left( \Spin^{(1)}_{n} .
\Spin^{(2)}_{n+1}+ \Spin^{(1)}_{n+1} . \Spin^{(2)}_{n} \right) \ .
\eeq{crossH} where the last term corresponds to additional
diagonal couplings.

Following the same approach as for the normal ladder we introduce
Hubbard-Stratonovich fields associated to each coupling and
perform a three-parameter MF analysis proposing constant values
for the intrachain couplings, the interchain (rung) coupling and
the interchain (diagonal) couplings. We find two different regions
in the parameter space $(J/J',J_{\times}/J')$.
It should be noted that this Hamiltonian is
dual under the interchange $J\leftrightarrow J_{\times}$, then it
is enough to study the region $J_{\times} \le J$.

\vspace{.5cm}

\noindent (a) If $J/J'> \pi^2/8$, the MF analysis yields the
system in a weak coupling regime, and following all the same steps
as before, we arrive at the same effective Field Theory with the
noticeable change that the coefficient of the relevant
perturbations is now shifted as $J'- 2 J_\times$. As in the weak
coupling analysis \cite{WKO,AEN}, one immediately sees that there is a
line in which the relevant perturbations vanish.
On this line one could expect a
massless regime, as suggested by numerical studies
\cite{WKO,W}. However in a recent treatment of the resulting
bosonized Hamiltonian it was shown that the current-current terms
are marginally relevant and a gap opens \cite{AEN}. The same
conclusion is attained in our resulting effective theory.
Again, the new feature here is that we find the
region of validity of the weak coupling effective Field Theory to
go up to $J' = 8/\pi^2 J$.

\vspace{.5cm}

\noindent (b) If $J/J' < \pi^2/8$, the system falls in a strong
coupling regime in which the two dispersion bands are separated by
a gap ($\propto J'$) and then a low-energy effective Field Theory
description is not suitable here.

\begin{figure}[hpt]
\begin{center}
\epsfig{file=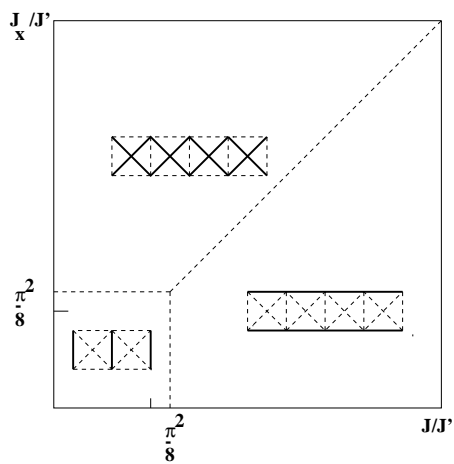,width=6cm,angle=-90}
\end{center}
\vskip .5 truecm

\caption{Phase diagram of the crossed ladder. Bold bonds
correspond again to non zero links in the MF approximation.}
\end{figure}

\vspace{.5cm}

\noindent {\it ii) Zig-zag ladders}

\vspace{.5cm}

The Hamiltonian is given by

\beq H=\sum_{n=1}^N  \sum_{l=1}^2 J \Spin^{(l)}_{n} .
\Spin^{(l)}_{n+1} + J' \sum_{n=1}^N \Spin^{(1)}_{n} . \left(
\Spin^{(2)}_{n}+ \Spin^{(2)}_{n+1} \right)\ .
\eeq{zzH}
Introducing again Hubbard-Stratonovich fields associated to each
coupling and performing a MF analysis with constant values for the
intrachain and interchain couplings, we find a different
situation: while we still find a regime, which now exists for $J'
< J$, in which we re-obtain the standard weak coupling results, we
find that the ``strong coupling" regime, ($J'
> J$), can still be described by an effective low energy Field
Theory.

More precisely, in the regime in which $J' < J$ we find that all
relevant perturbations cancel in a way similar to that found in
the weak coupling limit \cite{WA,AS,NGE,CHP}. The effective low
energy theory corresponds to the same coset theory, perturbed only
by the operators appearing in the first, third and fourth lines in
eq.\ (\ref{merda}). The so-called parity breaking terms first
studied in \cite{NGE} appear in the present approach from the
next-to-leading order in the lattice-spacing $a$ in the expansion
of the modified version of (\ref{PF}).

In the other regime, ($J' > J$), the bands at the MF minimum are
given by

\beqn \epsilon^{(+)}(k) &=& - \frac{J'}{\pi}\sqrt{2(1+\cos(k a))},
\nn \\ \epsilon^{(-)}(k) &=&  \frac{J'}{\pi}\sqrt{2(1+\cos(k a))}\
,
\eeqn{bandszz}
being no gap between them, and a Field Theory description is still possible.
The difference is that the low energy effective theory should in
this case be built up on only two fermion species, exhibiting
$SU(2)_1$ symmetry. This should correspond to the description of a
single chain plus next-nearest-neighbour interactions, which is
the suitable picture for the regime where $J'$ dominates.

\begin{figure}[hpt]
\begin{center}
\epsfig{file=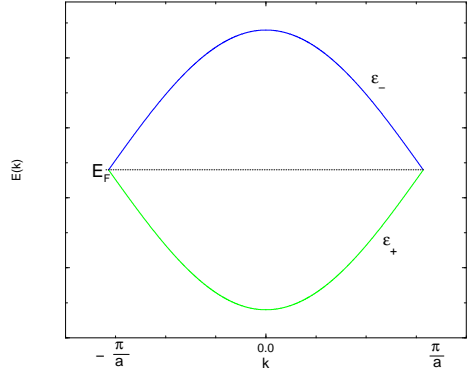,width=6cm,angle=-90}
\end{center}
\vskip .5 truecm

\caption{Energy bands for the zig-zag ladder at $J' > J$ MF minimum }
\end{figure}

Once
again, our method allows for the construction of an effective
field Theory for the full range of couplings and in particular
would allow to study the transition from the massless ($c=1$)
$J=0$ limit to the massive Kosterlitz-Thouless regime known to
arise at $J \approx 0.24 J'$ \cite{E}, which should, according to
our analysis, extend to the limit $J' \rightarrow 0$.

\begin{figure}[hpt]
\begin{center}
\epsfig{file=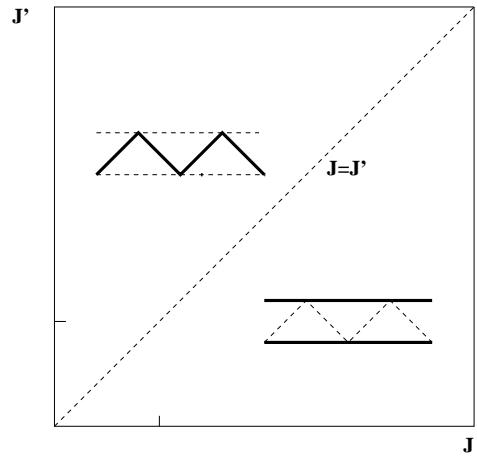,width=6cm,angle=-90}
\end{center}
\vskip .5 truecm
\caption{Phase diagram of the zig-zag ladder. Bold bonds
correspond again to non zero links in the MF approximation.}
\end{figure}

Since the main purpose of the present paper is to emphasize the
potential applications of our approach, the analysis of these
effective field theories will be addressed in a separate
publication.

\section{Conclusions}

The approach developed in the present paper shows that the spectrum
predicted from weak coupling approximation extends up to a finite
value of $J'/J$, our estimation of this  critical value being
$\frac{J'_c}{J}\sim \frac{8}{\pi^2}$. Beyond this value our MF
analysis of Section II predicts a cross-over to the strong coupling regime, where the
rungs of the ladder become disconnected among them. Fluctuations
over this state will restore connectivity and the strong coupling
approach of \cite{Sachdev,Gopalan} would be the appropriate
starting point in this parameter regime.
As the  classical
potential analyzed in Section II has a double well structure in
the intermediate region ($J'\sim J'_c$) we expect a smooth
cross-over from weak to strong coupling regime.
Experimental
observation of this cross-over supposes the variation of the ratio
of the exchange parameter. This could in principle be achieved by
applying pressure in the perpendicular direction of the ladder
axis.

Though our approach starts from a MF analysis, fluctuations are
taken into account to all orders. Besides, it allows for a
classification of all the perturbations in the language of the
embedding of the theory into $SU(2)_2 \times Z_2$.
One interesting observation which arises is that only the $Z_2$ Majorana
Fermi velocity is renormalized to first order in $J'$ by the
interacions.

The study of hole doped spin ladders is a natural extension of our
approach. For this case the $t-J$ model should be considered and
the charge sector of the theory could be represented by a spinless
boson (the slave boson representation). However the magnetic
excitations will evolve from the triplet and the singlet found in
this paper. The question of the hole pairing due to these
excitations could therefore be addressed within our formalism.
This will be reported elsewhere.

{\em Acknowledgements}: We are grateful to E.F.\ Moreno for useful
discussions and computational help. We thank A. Greco, A.\
Honecker, A.A.\ Nersesyan and P.\ Pujol for useful comments. We
thank CONICET and Fundaci\'on Antorchas (grants No.\ A-13622/1-106
and A-13740/1-64) for financial support.

\section*{Appendix}

We write in this appendix the explicit form of some lengthy
expressions appearing with compact notation in the main text.

The relevant part of the third term in eq.\ (\ref{PF}), appearing
in eq.\ (\ref{coset}), reads in the continuum limit
\beqn
\lefteqn { \mbox{relevant perturbations} =}\nonumber \\
&-& \frac{J'a}{2} \int dx \left( -\Psi^{\dagger (-)}_{R, \alpha}\Psi^{\dagger
(-)}_{R, \beta} \Psi^{(-)} _{L, \alpha}\Psi^{(-)} _{L, \beta} +
\Psi^{\dagger (-)}_{R, \alpha}\Psi^{\dagger (-)}_{R, \beta}
\Psi^{(-)} _{L, \alpha}\Psi^{(+)} _{L, \beta} - \Psi^{\dagger
(-)}_{R, \alpha}\Psi^{\dagger (-)}_{R, \beta} \Psi^{(+)} _{L,
\alpha}\Psi^{(-)} _{L, \beta}+ \Psi^{\dagger (-)}_{R,
\alpha}\Psi^{\dagger (-)}_{R, \beta} \Psi^{(+)} _{L,
\alpha}\Psi^{(+)} _{L, \beta}\right. \nonumber \\ & -&
\Psi^{\dagger (-)}_{R, \alpha}\Psi^{\dagger (+)}_{R, \beta}
\Psi^{(-)} _{L, \alpha}\Psi^{(-)} _{L, \beta }+ \Psi^{\dagger
(-)}_{R, \alpha}\Psi^{\dagger (+)}_{R, \beta} \Psi^{(-)} _{L,
\alpha}\Psi^{(+)} _{L, \beta} - \Psi^{\dagger (-)}_{R,
\alpha}\Psi^{\dagger (+)}_{R, \beta} \Psi^{(+)} _{L,
\alpha}\Psi^{(-)} _{L, \beta} + \Psi^{\dagger (-)}_{R,
\alpha}\Psi^{\dagger (+)}_{R, \beta} \Psi^{(+)} _{L,
\alpha}\Psi^{(+)} _{L, \beta}  \nonumber \\ &+ & \Psi^{\dagger
(+)}_{R, \alpha}\Psi^{\dagger (-)}_{R, \beta} \Psi^{(-)} _{L,
\alpha}\Psi^{(-)} _{L, \beta} - \Psi^{\dagger (+)}_{R,
\alpha}\Psi^{\dagger (-)}_{R, \beta} \Psi^{(-)} _{L,
\alpha}\Psi^{(+)} _{L, \beta} + \Psi^{\dagger (+)}_{R,
\alpha}\Psi^{\dagger (-)}_{R, \beta} \Psi^{(+)} _{L,
\alpha}\Psi^{(-)} _{L, \beta} - \Psi^{\dagger (+)}_{R,
\alpha}\Psi^{\dagger (-)}_{R, \beta} \Psi^{(+)} _{L,
\alpha}\Psi^{(+)} _{L, \beta} \nonumber \\ &+& \left.
\Psi^{\dagger (+)}_{R, \alpha}\Psi^{\dagger (+)}_{R, \beta}
\Psi^{(-)} _{L, \alpha}\Psi^{(-)} _{L, \beta} - \Psi^{\dagger
(+)}_{R, \alpha}\Psi^{\dagger (+)}_{R, \beta} \Psi^{(-)} _{L,
\alpha}\Psi^{(+)} _{L, \beta} + \Psi^{\dagger (+)}_{R,
\alpha}\Psi^{\dagger (+)}_{R, \beta} \Psi^{(+)} _{L,
\alpha}\Psi^{(-)} _{L, \beta} - \Psi^{\dagger (+)}_{R,
\alpha}\Psi^{\dagger (+)}_{R, \beta} \Psi^{(+)} _{L,
\alpha}\Psi^{(+)} _{L, \beta}
 +  H.c.\ \right)
\eeqn{PRfermiones} by simple use of eqs.\ (\ref{psi+}),
(\ref{psi-}), (\ref{z123}), (\ref{wv}).

The explicit form of eq.\ (\ref{asym}) in terms of fermions, using
eq.\ (\ref{bosrule}), is
\beq \Phi_{\stackrel{\textstyle{\small\Box}}{\Box
}}^{[(i_1\alpha_1),(i_2,\alpha_2)],
[(\bar\jmath_1\bar\beta_1),(\bar\jmath_2,\bar\beta_2)] } = {\cal
A} \left( \Phi_{\Box}^{(i_1\alpha_1),(\bar\jmath_1\bar\beta_1)}
\Phi_{\Box}^{(i_2,\alpha_2),(\bar\jmath_2,\bar\beta_2)} \right)\ ,
\eeq{expasym} where antisymmetrization affects bar and unbar pairs
of indices separately.

Using eqs.\ (\ref{PRfermiones}) and (\ref{expasym}), expression
(\ref{pertfull}) follows immediately. The base used for writing
the matrix $A$ in eq.\ (\ref{matrix}) is the one made explicit with
indices in the l.h.s.\ of eq.\ (\ref{sim}), ordered as
$ [(+,\uparrow),(-,\uparrow)],[(+,\uparrow),(+,\downarrow)],[(+,\uparrow),(-,\downarrow)],
[(-,\uparrow),(+,\downarrow)],[(-,\uparrow),(-,\downarrow)],
[(+,\downarrow),(-,\downarrow)]$.

\end{document}